\documentclass[aps,prl,letterpaper,twocolumn,showpacs,,amsmath,amssymb,superscriptaddress]{revtex4-1}
\usepackage{amsmath}
\usepackage{graphicx}
\usepackage{setspace}
\usepackage{amssymb}
\usepackage{dcolumn}
\usepackage{mathrsfs}
\usepackage{bm}

\begin{document}

\preprint{}

\title{Optimal Storage and Retrieval of Single-Photon Waveforms}

\author{Shuyu Zhou}
\affiliation{Department of Physics, The Hong Kong University of Science and Technology, Clear Water Bay, Kowloon, Hong Kong, China}

\author{Shanchao Zhang}
\affiliation{Department of Physics, The Hong Kong University of Science and Technology, Clear Water Bay, Kowloon, Hong Kong, China}

\author{Chang Liu}
\affiliation{Department of Physics, The Hong Kong University of Science and Technology, Clear Water Bay, Kowloon, Hong Kong, China}

\author{J. F. Chen}
\affiliation{Department of Physics, The Hong Kong University of Science and Technology, Clear Water Bay, Kowloon, Hong Kong, China}

\author{Jianming Wen}
\affiliation{Institute for Quantum Information Science and Department of Physics and Astronomy, University of Calgary, Calgary T2N 1N4, Alberta, Canada}

\author{M. M. T. Loy}
\affiliation{Department of Physics, The Hong Kong University of Science and Technology, Clear Water Bay, Kowloon, Hong Kong, China}

\author{G. K. L. Wong}
\affiliation{Department of Physics, The Hong Kong University of Science and Technology, Clear Water Bay, Kowloon, Hong Kong, China}

\author{Shengwang Du} \email{dusw@ust.hk}
\affiliation{Department of Physics, The Hong Kong University of Science and Technology, Clear Water Bay, Kowloon, Hong Kong, China}
\date{\today}

\begin{abstract}
We report an experimental demonstration of optimal storage and retrieval of heralded single-photon wave packets using electromagnetically induced transparency (EIT) in cold atoms at a high optical depth. We obtain an optimal storage efficiency of (49$\pm$3)\% for single-photon waveforms with a temporal likeness of 96\%. Our result brings the EIT quantum light-matter interface close to practical quantum information applications.
\end{abstract}


\maketitle

Storage and retrieval of single photons with preserved quantum states is of great importance for long-distance quantum communication and quantum computation \cite{nphoton3-706, SimonPRL2007, DLCZ}. A practical quantum memory is desirable with high storage efficiency, long coherence time, and low noise. In the past decade, many schemes have been proposed and demonstrated for optical storage based on coherent light-matter interactions, such as electromagnetically induced transparency (EIT) \cite{EITHarris, MFleischhauer, HauNature2001}, off-resonance Raman interaction \cite{WalmsleyPRL2011}, and photon echo \cite{Alexander}. Of these techniques, photon echo has recently become attractive due to its promising storage efficiency (as high as 87\%), large mode capacity, and compatibility with solid state interfaces \cite{BCBucler2011, nature465-1052, Riedmatten2008}. However, these experimental demonstrations of high efficiencies were all limited to coherent light pulses, and the recent implementation with entangled photons only achieved a memory efficiency of about 2\% \cite{TittelNature2011, TittelPRL2012}.

On the other hand, the EIT memory is compatible with quantum state operation of single-photon wave packets \cite{WenPRA2004, Kuzmich2005,MDLuking2005,nature452-67} and squeezed states \cite{MKozuma,AILvovsky}. Recent progress includes storing narrow-band single photons generated from atomic systems \cite{Kuzmich2005,MDLuking2005,nature452-67} and spontaneous parametric down conversion \cite{PanNP2011}. Also, the EIT memory time has been pushed to milliseconds by prolonging the ground-state coherence \cite{PanNP2009}. However, EIT quantum memories have suffered from low efficiency so far, with the highest single-photon storage efficiency being only 17\% \cite{nature452-67, RempePRL2011}, preventing the scheme from practical applications.

In this Letter, we report an experimental demonstration of efficient storage and retrieval of narrow-band single-photon waveforms using EIT in a cold atomic ensemble. With the ability to control both single-photon wave packets and the memory bandwidth, we obtain a storage efficiency up to (49$\pm$3)\% while the nonclassical property is maintained. To our knowledge, it represents the highest storage efficiency for a single-photon waveform to date. Because an efficiency above 50\% is necessary to operate a memory within the non-cloning regime and beat the classical limit \cite{GrangierPRA2001}, our result brings the atomic quantum light-matter interface closer to practical quantum information applications \cite{nature465-1052}.

\begin{figure}
\includegraphics[width=7.5cm]{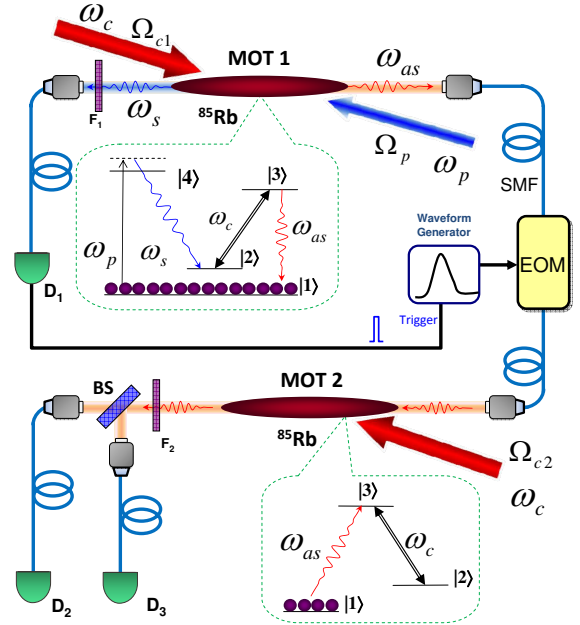}
\caption{\label{fig:systemconfiguration}(color online). Schematics of the experimental setup for storage and retrieval of heralded single photons with controllable waveforms. The $^{85}$Rb energy levels are chosen as $|1\rangle=|5S_{1/2},F=2\rangle$, $|2\rangle=|5S_{1/2},F=3\rangle$, $|3\rangle=|5P_{1/2},F=3\rangle$ and $|4\rangle=|5P_{3/2},F=3\rangle$.}
\end{figure}

Figure \ref{fig:systemconfiguration} illustrates the experimental configuration we use to generate, store, and retrieve narrow-band single photons. We make use of two two-dimensional (2D) $^{85}$Rb magneto-optical traps (MOT1 and MOT2), similar to the setup described in our previous work \cite{SinglePhotonPrecursor}. Each cold atomic cloud, with a temperature of about 100 $\mu$K, has a length of 1.7 cm and transverse diameter of 0.7 mm. From MOT1, we produce Stokes ($\omega_s$) and anti-Stokes ($\omega_{as}$) paired photons \cite{prl100-183603}, with the presence of counter-propagating pump ($\omega_p$, 780nm) and coupling ($\omega_c$, 795nm) beams aligned at a $3^\circ$ angle with respect to the Stokes-anti-Stokes axis. The pump laser is blue detuned by 60 MHz from the $|1\rangle \rightarrow |4\rangle$ transition. The coupling laser is on resonance with the $|2\rangle \rightarrow |3\rangle$ transition. Both the pump and coupling lasers have the same collimated beam diameter of 1.6 mm and their linewidths are narrower than 1 MHz. The Stokes and anti-Stokes photons are coupled into two opposing single-mode fibers (SMF). When the Stokes photon is detected by the single-photon detector $D_1$, we send its paired anti-Stokes photon through an amplitude electro-optical modulator (EOM, 10 GHz, EOspace), which is driven by a triggered waveform generator. In this way, we are able to generate heralded single anti-Stokes photons with controllable waveforms \cite{EOM}. We then store the anti-Stokes photons in the cold atoms at MOT2, controlled by a second coupling beam directed from the same coupling laser in MOT1. The anti-Stokes photon single mode is focused to the center of MOT2 along its longitudinal axis and has a $1/e^2$ diameter of 245 $\mu$m at the waist. The coupling beam at MOT2, with a $1/e^2$ diameter of 1.0 mm, is aligned at a $3^\circ$ angle with respect to the anti-Stokes propagation. We run the experiment periodically with a MOT time of 4.5 ms followed by a photon generation window of 0.5 ms for each cycle. The MOT magnetic fields remain on all the time. In both MOTs, at end of the trapping time, we optically pump all the atoms to the ground level $|1\rangle$. Coincidence counts are recorded by a time-to-digital converter (Fast Comtec P7888) with 1 ns bin width.

The physical mechanism of EIT memory has been well studied in terms of dark-state polaritons \cite{MFleischhauer}. As the photon wave packet is spatially compressed inside the medium, we turn off the coupling laser to adiabatically convert the photon state into a long-lived atomic spin wave that involves only the two ground levels $|1\rangle$ and $|2\rangle$. After a controllable time delay, we turn on the coupling laser again to retrieve the photon wave packet. There are two important parameters characterizing the performance of a single-photon memory. The first is the storage efficiency, defined as the probability of storing and retrieving the single photon,
\begin{eqnarray}
\eta=\frac{\int|\psi_{out}(\tau)|^2d\tau}{\int|\psi_{in}(\tau)|^2d\tau}, \label{eq:storageEfficiency}
\end{eqnarray}
where $\psi_{in}(\tau)$ and $\psi_{out}(\tau)$ are the input and output heralded single-photon wave packets with $\tau=t_{as}-t_s$. The storage efficiency is determined by both the photon temporal waveform and the EIT memory bandwidth. The second parameter is the storage time, which is limited by the ground-state coherence time. In this work, we focus on the storage efficiency at two pulse-length storage time. Moreover, a single photon storage requires the memory to be operated at an ultra-low noise level. For the EIT memory, the major noise comes from the scattering of the coupling laser beam. Compared to warm atomic vapor cells that require a collinear Doppler-free optical setup \cite{MDLuking2005}, this scattering is suppressed in our cold atom system because of the $3^\circ$ angle between the coupling beam and the anti-Stokes photons. Further noise reduction is accomplished by two optical frequency filters ($F_1$ and $F_2$, with a bandwidth of 0.5 GHz). A beam splitter (BS) and two detectors ($D_2$ and $D_3$) are used to verify the single-photon quantum nature, because a single photon incident at a BS must go to one port or the other. A measure of the quality of heralded single photons is given by the conditional correlation function \cite{ConditionalG2}
\begin{eqnarray}
g_{c}^{(2)}=\frac{N_{123}N_{1}}{N_{12}N_{13}}, \label{eq:gc2}
\end{eqnarray}
where $N_1$ is the Stokes counts at $D_1$, $N_{12}$ and $N_{13}$ are the twofold coincidence counts, and $N_{123}$ is the threefold coincidence counts. A classical field must satisfy $g_{c}^{(2)}\geq 1$. A pure single photon has $g_{c}^{(2)}=0$ and a two-photon state has $g_{c}^{(2)}=0.5$. Therefore $g_{c}^{(2)}<1.0$ violates the classical limit and $g_{c}^{(2)}<0.5$ suggests the near-single-photon character.

\begin{figure}
\includegraphics[width=7.5cm]{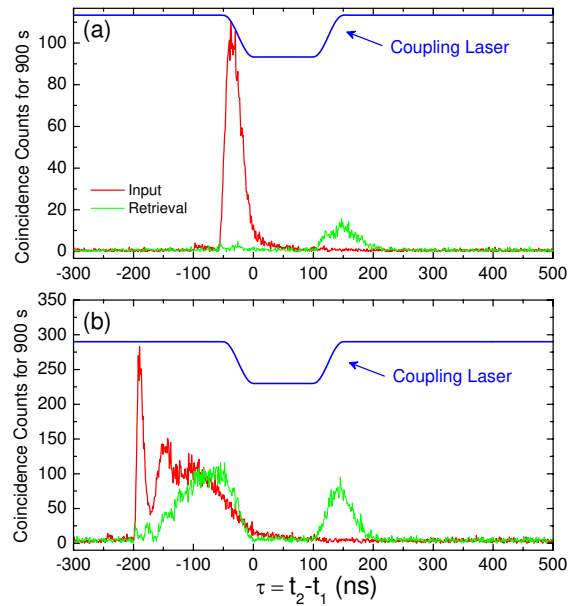}
\caption{\label{fig:originalstorage}(color online). Direct storage and retrieval of single photons without waveform shaping. Single photons with (a) a short waveform and (b) a long waveform are produced from MOT1 at OD$_1$=7 and 35, respectively. The coincidence counts are recorded by D$_1$ and D$_2$. Other parameters are OD$_2$=60, $\Omega_{c2}=11\gamma_{13}$, and $\gamma_{12}=0.03\gamma_{13}$.}
\end{figure}

We first characterize the photon source. In the following experiments, we fix the pump and coupling laser Rabi frequencies during the biphoton generation in MOT1 at $\Omega_p=0.4 \gamma_{13}$ and $\Omega_{c1}=5.1 \gamma_{13}$, where $\gamma_{13}=2\pi\times3$ MHz is the electric dipole relaxation rate between $|1\rangle$ and $|3\rangle$. The optical depth of MOT2 is maintained at OD$_2$=60. By varying the OD at MOT1 (OD$_1$), we produce paired photons with controllable temporal length \cite{prl100-183603}. The red curves in Fig.~\ref{fig:originalstorage} show the two-photon coincidence counts between detectors D$_1$ and D$_2$ with 1 ns bin width, collected for 900 s. At OD$_1$=7, the heralded anti-Stokes photon has a temporal length of about 50 ns, as shown in Fig.~\ref{fig:originalstorage}(a), while at OD$_1$=35 we prolong the length to 200 ns as shown in Fig.~\ref{fig:originalstorage}(b). Excluding the uncorrelated accidental coincidences, there are a total of 3300 (or 18100) biphoton coincidence counts detected by D$_1$ and D$_2$ for OD$_1$=7 (or 35). Including the coincidence counts between D$_1$ and D$_3$ (the measured BS splitting ratio is about 45\%:55\%), we detect a total of 7400 (or 42400) photon pairs in 900 s, corresponding to a photon pair detection rate of 8 (or 47) pair/s. Taking into account the detector quantum efficiencies (50\% each), fiber-fiber coupling efficiencies (70\% at MOT1 and 72\% at MOT2), EOM transmission (50\%), fiber connection efficiency (61\%), filter transmissions (65\% each), and the duty cycle (10\%), this corresponds to a generation rate of about 4900 (or 28900) pair/s from MOT1. At OD$_1$=7 (or 35), for each click at D$_1$, the success probability of detecting its heralded photon at D$_2$ and D$_3$ is 2.8\% (or 4.1\%), which, accounting all the losses and efficiencies, corresponds to a pairing efficiency of 56\% (or 82\%) when they are produced from MOT1. The incident anti-Stokes photon rate in MOT2 is about 1000/s (or 6200/s). The nonclassical properties of the paired photons can be measured by violation of the Cauchy-Schwartz inequality $[g^{(2)}_{i,j}(\tau)]^2/[g^{(2)}_{i,i}(0)g^{(2)}_{j,j}(0)]\leq 1$ \cite{Clauser}. Because the paired photons are generated through spontaneous four-wave mixing, there is no correlation between different pairs. Therefore, the correlation function $g^{(2)}_{s,as}(\tau)$ can be obtained by normalizing the two-photon coincidence counts to the background floor resulting from accidental coincidences between uncorrelated photons. We obtain $g^{(2)}_{s,as}(\tau)$ with maximum values of 150 and 95 for the input waveforms in Fig.~\ref{fig:originalstorage}(a) and (b), respectively. With $g^{(2)}_{s,s}(0)=g^{(2)}_{as,as}(0)=2.0$, we obtain a violation of the inequality by a factor of 5625 and 2256, respectively. To characterize the single-photon nature of the heralded photons, we measure $g_{c}^{(2)}=0.10\pm0.02$ for the short photon (with a coincidence window of 100 ns ) and $g_{c}^{(2)}=0.17\pm0.02$ for the long photon (with a coincidence window of 200 ns ), each with a total time of 2100 s.

We then measure the storage efficiency without shaping the waveform of anti-Stokes photons by leaving the EOM at its maximum transmission. To store the anti-Stokes photon, we switch off the coupling laser at MOT2 ($\Omega_{c2}=11 \gamma_{13}$) for a period of 100 ns after detecting its paired Stokes photon. The retrieved photon waveforms are displayed as the green curves in Fig.~\ref{fig:originalstorage}(a) and (b). The coupling laser has switch-on and -off times of 50 ns. For both waveforms, we obtain the same storage efficiency of (20$\pm$2)\%. The measured $g_{c}^{(2)}=0.24\pm0.17$ and $0.44\pm0.15$ confirm that we indeed retrieve single photons. However, in both cases, the waveform profiles are not preserved after retrieval.

\begin{figure}
\includegraphics[width=7.5cm]{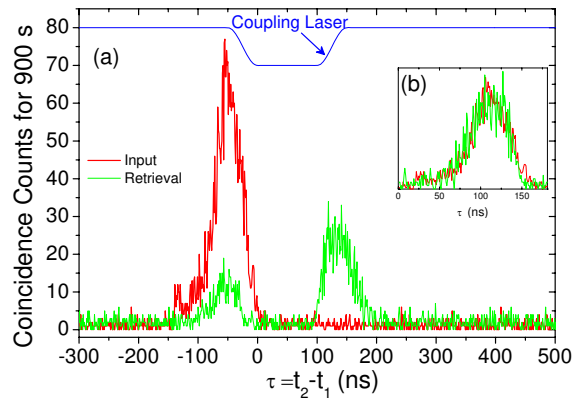}
\caption{\label{fig:optstorage}(color online). Storage and retrieval of a single photon with optimal waveform. (a) The optimal input (red curve) and output (retrieval, green curve) heralded single-photon waveforms are measured as coincidence counts between D$_1$ and D$_2$. (b) The inset shows the time-reversed retrieved photon waveform matches the input photon waveform after normalization. The operating parameters at MOT2 are OD$_2$=60, $\Omega_{c2}=11\gamma_{13}$, and $\gamma_{12}=0.03\gamma_{13}$. The measured storage efficiency is (36$\pm$3)\%.}
\end{figure}

Previous work for coherent pulse storage suggests that the storage efficiency can be substantially improved by optimizing the pulse shape to match the EIT bandwidth. We follow the optimization procedure described in \cite{optcontrol2007}. We work with the initial waveform in Fig.~\ref{fig:originalstorage}(b). The oscillatory structure on top of the leading edge is an optical precursor at the single-photon level \cite{SinglePhotonPrecursor, prl100-183603, BiphotonPrecursor}, which is produced from the optical frequency components far away from the atomic resonance. This optical precursor component can not be slowed and stored \cite{SinglePhotonPrecursor}, thus we remove it from the single-photon waveform using the EOM. At MOT2, we fix the coupling laser Rabi frequency as $\Omega_{c2}=11 \gamma_{13}$ and OD$_2$=60. We feed back the time-reversed waveform of the retrieved photon to shape its input waveform using the EOM. The optimal input-output waveforms are obtained after three iterations, as shown in Fig.~\ref{fig:optstorage}(a).  The input single-photon waveform has a full width at half maximum (FWHM) of 50 ns and a peak $g^{(2)}_{s,as}$ of 23. The output single-photon waveform is retrieved after a storage time of 2 pulse length (100 ns). The optimal storage efficiency is (36$\pm$3)\%. To confirm that at this optimal condition, the retrieved photon waveform is the time-reversal of the input, we plot the time-reversed retrieved photon waveform together with the input waveform with a proper rescaling in the inset [Fig.~\ref{fig:optstorage}(b)], and they match each other very well. The result agrees with the theoretical prediction and indicates that the amplitude-phase information of the single-photon wave packet is preserved during storage \cite{opttheo2007}. For a quantitative estimation, we calculate the temporal waveform likeness $L=|\int \psi_{in}(\tau)\psi_{out}(-\tau)d\tau|^2/[\int |\psi_{in}(\tau)|^2 d\tau \int |\psi_{out}(\tau)|^2 d\tau]$. Since we work at the group delay regime, we obtain transform-limited single-photon waveforms after amplitude modulation \cite{SinglePhotonPrecursor, DuJOSAB2008}. Therefore, we have $\psi_{in}(\tau)=\sqrt{G_{in}^{(2)}(\tau)}$ and $\psi_{out}(\tau)=\sqrt{G_{out}^{(2)}(\tau)}$, where $G_{in}^{(2)}(\tau)$ and $G_{out}^{(2)}(\tau)$ are Glauber correlation functions before and after storage that can be obtained from the coincidence counts in Fig.~\ref{fig:optstorage}. For this optimal storage, we obtain $L$=93\%. We measure $g_c^{(2)}=0.30\pm0.06$ for the input photon and $g_c^{(2)}=0.26\pm0.13$ for the retrieved photon with a coincidence window of 100 ns and a total time of 3300 s. They are below the two-photon threshold of 0.5.

\begin{figure}
\includegraphics[width=7.5cm]{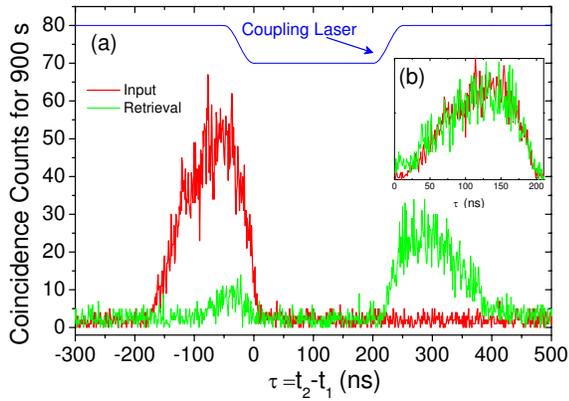}
\caption{\label{fig:higheffstorage}(color online). Optimal storage and retrieval of single photons with reduced ground-state dephasing rate. (a) The optimal input (red curve) and output (retrieval, green curve) heralded single-photon waveforms. (b) The inset shows the time-reversed retrieved photon waveform matches the input photon waveform after normalization. The operating parameters at MOT2 are OD$_2$=60, $\Omega_{c2}=6.88\gamma_{13}$, and $\gamma_{12}=0.01\gamma_{13}$. The measured storage efficiency is (49$\pm$3)\%.}
\end{figure}

The storage efficiency is also affected by the dephasing rate $\gamma_{12}$ between the two ground levels $|1\rangle$ and $|2\rangle$ which causes decoherence loss and absorption at high OD. We obtain the effective dephasing rate by best-fitting it to the EIT transmission spectrum. The finite dephasing rate is caused by stray magnetic fields, atomic motion, and the coupling beam profile. In the above measurements, the coupling beam at MOT2 has a diameter of 1.0 mm, and we obtain $\gamma_{12}=0.03\gamma_{13}$. To reduce the dephasing rate, we increase the coupling beam diameter to 1.6 mm and obtain $\gamma_{12}=0.01\gamma_{13}$. The coupling laser power, which is limited by the maximum power allowed by the fiber EOM, remains the same as for the previous measurements and thus we have $\Omega_{c2}= 6.88 \gamma_{13}$. Figure \ref{fig:higheffstorage}(a) shows the optimal storage and retrieval of single photon waveforms under the new conditions. Due to the narrower EIT bandwidth, the optimal waveform has a longer temporal length (FWHM=100 ns) and a peak $g^{(2)}_{s,as}$ of 12. For a delay of two pulse lengths (200 ns), we obtain a storage efficiency of (49$\pm$3)\%. The inset shows that the suitably rescaled time-reversed output waveform matches the input waveform well [Fig.~\ref{fig:higheffstorage}(b)]. We obtain a temporal waveform likeness of $L$=96\%. For the conditional correlation measurement, we take the FWHM (100 ns) as the coincidence window and obtain $g_c^{(2)}=0.10\pm0.06$ for the input waveform and $g_c^{(2)}=0.14\pm0.14$ for the retrieved photon. The higher $g_c^{(2)}$ after retrieval is caused by the single photon storage loss in presence of detector dark counts and multiphoton events from the background noise photons. As we increase the coincidence window to 200 ns to contain more accidental counts from noise photons and dark counts, the $g_c ^{(2)}$ of the input and retrieved photon become $0.28\pm0.07$ and $0.66\pm0.22$ respectively, which are still below the classical limit. The measured memory lifetime is about 1.6 $\mu$s, which is mainly determined by the inhomogeneous MOT magnetic field.

In our configuration, we find that the connection between $g_c^{(2)}$ and the normalized cross correlation function $\bar{g}_{s,as}^{(2)}$ (averaged over the same coincidence window) can be expressed as $g_c^{(2)}\simeq (2 \bar{g}_{s,as}^{(2)}+1)/[(\bar{g}_{s,as}^{(2)}+1)^2]$. In the case $\bar{g}_{s,as}^{(2)}>>1$, this reduces to $g_c^{(2)}\simeq 2/\bar{g}_{s,as}^{(2)}$, which agrees with our experimentally measured values within their statistical errors. As we increase the coincidence window length, $g_c^{(2)}$ increases because of an increasing probability for detecting multiphoton events from uncorrelated noise photons and dark counts. Oppositely, $\bar{g}_{s,as}^{(2)}$ drops as the coincidence window length increases. As a measure of the ratio of correlated photons (signal) to uncorrelated photons (noise), $\bar{g}_{s,as}^{(2)}$ provides a quick estimate of the quality of heralded single photons. It is clear that both $g_c^{(2)}<1$ and the violation of Cauchy-Schwartz inequality require  $\bar{g}_{s,as}^{(2)}>2$ for beating the classical limit. $\bar{g}_{s,as}^{(2)}>4$ indicates the near-single-photon character of the heralded anti-Stokes photon ($g_c^{(2)}<0.5$).

In summary, we have demonstrated optimal storage and retrieval of heralded single-photon wave packets using EIT in cold atoms. At a high OD of 60, we obtain a storage efficiency of close to 50\% for the optimal single-photon waveform with a temporal likeness of 96\%. The storage efficiency in this work refer only to the capability of the EIT atomic medium at MOT2 in storing and retrieving optimal single-photon waveforms, and it does not include the fiber connection loss and the EOM insertion loss. The EOM amplitude modulation loss of about 50\% in our system can be counted into the heralded single-photon generation efficiency. This modulation loss can be eliminated, in principle, using other waveform shaping techniques, such as chirp (using phase-frequency modulation) and compression \cite{HarrisPRL2007, HarrisPRL2010}. 

The work was supported by the Hong Kong Research Grants Council (Project No. 601411). J. W. was supported by an AI-TF New Faculty Grant and an NSERC Discovery Grant.



\begin{thebibliography}{99}

\bibitem{nphoton3-706} A. I. lvovsky, B. C. Sanders and W. Tittel, Nature Photonics \textbf{3}, 706 (2009).

\bibitem{SimonPRL2007} C. Simon, H. de Riedmatten, M. Afzelius, N. Sangouard, H. Zbinden, and N. Gisin, Phys. Rev. Lett. \textbf{98}, 190503 (2007).

\bibitem{DLCZ} L. -M. Duan, M. D. Lukin, J. I. Cirac, and P. Zoller, Nature \textbf{414}, 413 (2001).

\bibitem{EITHarris} S. E. Harris, Phys. Today \textbf{50}, 36 (1997).

\bibitem{MFleischhauer} M. Fleischhauer, A. Imamoglu, and J. P. Marangos, Rev. Mod. Phys. \textbf{77,} 633 (2005).

\bibitem{HauNature2001} C. Liu, Z. Dutton, C. H. Behroozi, and L. V. Hau, Nature \textbf{409}, 490 (2001).

\bibitem{WalmsleyPRL2011} K. F. Reim, P. Michelberger, K. C. Lee, J. Nunn, N. K. Langford, and I. A. Walmsley, Phys. Rev. Lett. \textbf{107}, 053603 (2011).

\bibitem{Alexander} A. L. Alexander, J. J. Longdell, M. J. Sellars, and N. B. Manson, Phys. Rev. Lett. \textbf{96}, 043602 (2006).

\bibitem{BCBucler2011} M. Hosseini, B. M. Sparkes, G. Campbell, P. K. Lam and B. C. Buchler, Nature Commun. \textbf{2}, 174 (2011).

\bibitem{nature465-1052} M. P. Hedges, J. J. Longdell, Y. Li, and M. J. Sellars, Nature \textbf{465}, 1052 (2010).

\bibitem{Riedmatten2008} H. de Riedmatten, M. Afzelius, M. U. Staudt, C. Simon, and N. Gisin, Nature \textbf{456}, 773 (2008).

\bibitem{TittelNature2011} E. Saglamyurek, N. Sinclair, J. Jin, J. A. Slater, D. Oblak, F. Bussi\`{e}res, M. George, R. Ricken, W. Sohler, and W. Tittel, Nature \textbf{469}, 512 (2011).
    
\bibitem{TittelPRL2012} E. Saglamyurek, N. Sinclair, J. Jin, J. A. Slater, D. Oblak, F. Bussi\`{e}res, M. George, R. Ricken, W. Sohler, and W. Tittel, Phys. Rev. Lett. \textbf{108}, 083602 (2012).

\bibitem{WenPRA2004} J. Wen and M. H. Rubin, Phys. Rev. A \textbf{70}, 063806 (2004).

\bibitem{Kuzmich2005} T. Chaneli\`{e}re, D. N. Matsukevich, S. D. Jenkins, S. -Y. Lan, T. A. B. Kennedy and A. Kuzmich, Nature \textbf{438}, 833 (2005).

\bibitem{MDLuking2005} M. D. Eisaman, A. Andr\'{e}, F. Massou, M. Fleischhauer, A. S. Zibrov and M. D. Lukin, Nature \textbf{438}, 837 (2005).

\bibitem{nature452-67} K. S. Choi, H. Deng, J. Laurat, and H. J. Kimble, Nature \textbf{452}, 67 (2008).

\bibitem{MKozuma} K. Honda, D. Akamatsu, M. Arikawa, Y. Yokoi, K. Akiba, S. Nagatsuka, T. Tanimura, A. Furusawa, and M. Kozuma, Phys. Rev. Lett. \textbf{100}, 093601 (2008).

\bibitem{AILvovsky} J. Appel, E. Figueroa, D. Korystov, M. Lobino, and A. I. Lvovsky, Phys. Rev. Lett. \textbf{100}, 093602 (2008).

\bibitem{PanNP2011} H. Zhang \emph{et al}., Nature Photon. \textbf{5}, 628 (2011).

\bibitem{PanNP2009} B. Zhao \emph{et al}., Nature Phys. \textbf{5}, 95 (2009).

\bibitem{RempePRL2011} M. Lettner, M. M\"{u}cke, S. Riedl, C. Vo, C. Hahn, S. Baur, J. Bochmann, S. Ritter, S. D\"{u}rr, and G. Rempe, Phys. Rev. Lett. 106, 210503 (2011).

\bibitem{GrangierPRA2001} F. Grosshans and P. Grangier, Phys. Rev. A \textbf{64}, 010301(R) (2001).

\bibitem{SinglePhotonPrecursor} S. Zhang, J. F. Chen, C. Liu, M. M. T. Loy, G. K. L. Wong, and S. Du, Phys. Rev. Lett. \textbf{106}, 243602 (2011).

\bibitem{prl100-183603} S. Du, P. Kolchin, C. Belthangady, G.Y. Yin, and S. E. Harris, Phys. Rev. Lett. \textbf{100}, 183603 (2008).

\bibitem{EOM} P. Kolchin, C. Belthangady, S. Du, G.Y. Yin, and S. E. Harris, Phys. Rev. Lett. \textbf{101}, 103601 (2008).

\bibitem{ConditionalG2} P. Grangier, G. Roger, and A. Aspect, Europhys. Lett. \textbf{1}, 173 (1986).

\bibitem{Clauser} J. F. Clauser, Phys. Rev. D \textbf{9}, 853 (1974).

\bibitem{optcontrol2007} I. Novikova, A. V. Gorshkov, D. F. Phillips, A. S. S{\o}rensen, M. D. Lukin, and R. L. Walsworth, Phys. Rev. Lett. \textbf{98}, 243602 (2007).


\bibitem{BiphotonPrecursor} S. Du, C. Belthangady, P. Kolchin, G. Y. Yin, and S. E. Harris, Opt. Lett. \textbf{33}, 2149 (2008).

\bibitem{opttheo2007} A. V. Gorshkov, A. Andr\'{e}, M. Fleischhauer, A. S. S{\o}rensen, and M. D. Lukin, Phys. Rev. Lett. \textbf{98}, 123601 (2007).

\bibitem{DuJOSAB2008} S. Du, J. Wen, and M. H. Rubin, J. Opt. Soc. Am. B \textbf{25}, C98 (2008).



\bibitem{HarrisPRL2007} S. E. Harris, Phys. Rev. Lett. \textbf{98}, 063602 (2007).

\bibitem{HarrisPRL2010} S. Sensarn, G. Y. Yin, and S. E. Harris, Phys. Rev. Lett. \textbf{104}, 253602 (2010).

\end{thebibliography}
\end{document}